\documentclass[journal]{IEEEtran}

\ifCLASSINFOpdf

\else

\fi
\usepackage{amsmath}
\usepackage{cite}

\usepackage{romannum}

\usepackage{amsmath,amsfonts,amssymb,enumerate,cases}
\usepackage{amsthm}
\usepackage{subfigure}
\usepackage{caption}
\usepackage[final]{lscape,graphicx,epsfig}
\usepackage{graphicx}
\usepackage{soul}
\usepackage[hmargin = 0.75in, vmargin=1.0in]{geometry}
\usepackage{xcolor}

\usepackage{bigints}
\usepackage{relsize}
\usepackage{lipsum}
\usepackage{changepage}

\usepackage{algorithm}
\usepackage[noend]{algpseudocode}
\usepackage{algpseudocode}
\makeatletter
\def\BState{\State\hskip-\ALG@thistlm}
\makeatother

\graphicspath{ {image/} }

\usepackage[T1]{fontenc}
\usepackage[english]{babel}
\usepackage{color}

\usepackage{algpseudocode}
\usepackage{algorithm}
\newcommand{\lyap}{M_t}

\newcommand{\Snoise}{\mathbf{W}}
\newcommand{\D}{\mathbf{D}}
\newcommand{\A}{\mathbf{A}}

\newcommand{\gra}{\mathcal{G}}
\newcommand{\ver}{\mathcal{V}}
\newcommand{\edg}{\mathcal{E}}

\newcommand{\pathset}{\mathcal{P}}
\newcommand{\neig}{\mathcal{N}}
\newcommand{\path}{p}
\usepackage{subfigure}
\newcommand{\exmax}{m_{+}}

\newtheorem{thm}{Theorem}
\newtheorem{lemma}{Lemma}

\usepackage{graphicx}

\usepackage{tikz}
\usetikzlibrary{matrix, positioning, calc, intersections, arrows, arrows.meta}
\tikzset{
	state/.style={shape=circle,draw=black!50,fill=blue!5},
	bits/.style={rectangle, text width=1.5em, align=center, text=white},
	greenthick/.style={-angle 60,very thick, green, line width=2pt},
	bluethick/.style={-angle 60,very thick, blue, line width=2pt},
	blueline/.style={blue, thick},
	redline/.style={red, thick},
}

\IEEEoverridecommandlockouts
\begin{document}
\title{Analysis and Design of Robust Max Consensus for Wireless Sensor Networks}
\author{Gowtham Muniraju, Cihan Tepedelenlioglu, \textit{Senior Member, IEEE} and Andreas Spanias, \textit{Fellow, IEEE}. \thanks{Parts
		of this paper were presented at the 2018 Asilomar Conference
		on Signals, Systems and Computers \cite{my_asilomar_max}.
		
		G. Muniraju, C. Tepedelenlioglu and A. Spanias are with the School of ECEE, Arizona State University. Email: \{gmuniraj,cihan,spanias\}@asu.edu.
		The authors from Arizona State University are funded in part by the NSF
		award ECSS 1307982, NSF CPS award 1646542 and the SenSIP Center,
		School of ECEE, Arizona State University.}\\
}
\date{}
\date{}
\include{macros_report}
\maketitle

\begin{abstract}
	A novel distributed algorithm for estimating the maximum of the node initial state values in a network, in the presence of additive communication noise is proposed. Conventionally, the maximum is estimated locally at each node by updating the node state value with the largest received measurements in every iteration. However, due to the additive channel noise, the estimate of the maximum at each node drifts at each iteration and this results in nodes diverging from the true max value. Max-plus algebra is used as a tool to study this ergodic process. The subadditive ergodic theorem is invoked to establish a constant growth rate for the state values due to noise, which is studied by analyzing the max-plus Lyapunov exponent of the product of noise matrices in a max-plus semiring. The growth rate of the state values is upper bounded by a constant which depends on the spectral radius of the network and the noise variance. Upper and lower bounds are derived for both fixed and random graphs. Finally, a two-run algorithm robust to additive noise in the network is proposed and its variance is analyzed using concentration inequalities. Simulation results supporting the theory are also presented. 
\end{abstract}
\begin{IEEEkeywords}
	Max consensus, spectral radius, max-plus algebra, wireless sensor networks.
\end{IEEEkeywords}

\IEEEpeerreviewmaketitle

\section{Introduction}
\label{sec1:intro}
A wireless sensor network (WSN) is a distributed network consisting of multi-functional sensors, which can communicate with neighboring sensors over wireless channels. 
Estimating the statistics of sensor measurements in WSNs is necessary in detecting anomalous sensors, supporting the nodes with insufficient resources, network area estimation \cite{app3_zhang2018distributed}, and spectrum sensing \cite{intro_spectrumsensing} for cognitive radio applications, just to name a few. Knowledge of extremes are often used in algorithms for outlier detection, clustering \cite{app1_muniraju2017location}, classification \cite{intro16_predd2006distributed}, and localization~\cite{app4_zhang2016node}.
However, several factors \cite{intro4_olfati2007consensus} such as additive noise in wireless channels, random link failures, packet loss and delay of arrival significantly degrade the performance of distributed algorithms. Hence it is important to design and analyze consensus algorithms robust to such adversities. 

Although max consensus has been studied in the literature \cite{a1_iutzeler2012analysis,r1_improved_maxbounds,max3_shi2012finite,a3_giannini2016asynchronous,a2_nejad2009max,aa1_nejad2010max}, the analysis of max consensus algorithms under additive channel noise and randomly changing network conditions has not received much attention. We start with a review of the literature on max consensus in the \textit{absence of noise}.
A distributed max consensus algorithm for both pairwise and broadcast communications is introduced in \cite{a1_iutzeler2012analysis} and also provides an upper bound on the mean convergence time. Recent work in \cite{r1_improved_maxbounds} consider pairwise and broadcast communications with asynchronous updates and significantly improve the tightness of the upper bound on the mean convergence time. 
 The convergence properties of max consensus protocols are studied in \cite{max3_shi2012finite,a3_giannini2016asynchronous,a2_nejad2009max,aa1_nejad2010max} for broadcast communications setting in distributed networks.
The convergence of average and max consensus algorithms in time dependent and state dependent graphs are analyzed in \cite{max3_shi2012finite}.
Asynchronous updates in the presence of bounded delays is considered in \cite{a3_giannini2016asynchronous}.  
Max-plus algebra is used to analyze convergence of max-consensus algorithms for time-invariant communication topologies in \cite{a2_nejad2009max}, and for switching topologies in \cite{aa1_nejad2010max}, both in the absence of noise. 
Distributed algorithms to reach consensus on general functions in the \textit{absence of noise} are studied in \cite{va1_tahbaz2006one,max1_cortes2008distributed,va2_bauso2006non}. 
A one-parameter family of consensus algorithms over a time-varying network is proposed in \cite{va1_tahbaz2006one}, where consensus on the minimum of the initial measurements can be reached by tuning a design parameter.
A distributed algorithm to reach consensus on general functions in a network is presented in \cite{max1_cortes2008distributed}, where the weighted power mean algorithm originally proposed by \cite{va2_bauso2006non} is used to calculate the maximum of the initial measurements by setting the design parameter to infinity.
 
A system model with imperfect transmissions is considered in \cite{r1_improved_maxbounds,aa1_nejad2010max}, where a message is received with a probability $1-p$. This model is equivalent to the time-varying graphs, where each edge is deleted independently with a probability $p$. 
However, these works do not consider errors in transmission, but only consider transmission failures (erasures).
 
Authors in \cite{a4_zhang2016max} considers the presence of additive noise in the network and propose an iterative soft-max based average consensus algorithm to approximate the maximum, which uses non linear bounded transmissions in order to achieve consensus. This algorithm depends on a design parameter that controls the trade-off between the max estimation error and convergence speed. However, the convergence speed of this soft-max based method is limited compared to the more natural max-based methods considered herein.

\subsection{Statement of Contributions} 
 The contribution of this paper is in both analysis of max consensus algorithms in presence of additive noise and design of fast max-based consensus algorithms. Due to additive noise, the estimate of the maximum at each node has a positive drift and this results in nodes diverging from the true max value. Max-plus algebra is used to represent this ergodic process of recursive max and addition operations on the state values. This growth rate is shown to be a constant in \cite{b1_van2007ergodic} for stochastic max-plus systems using the subadditive ergodic theorem, in a mathematics context that does not consider max-consensus. 
Even though the existence of growth rate follows from the sub-additive ergodic theorem, a formula on the rate itself is not available \cite{a5_heidergott2006max,b1_van2007ergodic}. In order to study the growth rate, we use large deviation theory and derive an upper bound for a general noise distribution in the network. We show that the upper bound depends linearly on the standard deviation, and is a function of the spectral radius of the network. Since the noise variance and spectral radius are not known locally at each node, we propose a two-run algorithm to locally estimate and compensate for the growth rate, and analyze its variance.

Our contributions beyond the conference version in \cite{my_asilomar_max}, are as follows. We include the complete proof of upper bound on the growth rate and also extend the analysis by deriving a lower bound. An empirical upper bound, which includes an additional correction factor that depends on number of nodes is shown to be tighter compared to \cite{my_asilomar_max}. Additionally, we derive the upper and lower bounds for time-varying random graphs, which model transmission failures, and additive noise. Furthermore, we present a method to directly calculate the upper bound, without solving for the large deviation rate function of the noise.
Also, using concentration inequalities we show that the variance of the growth rate estimator decreases inversely with the number of iterations and use this to bound the variance of our estimator. Through simulations, we show that our proposed algorithm converges much faster with lower estimation error, in comparison to existing algorithms. 

\subsection{Paper Organization}
The rest of this paper is organized as follows. The system model and problem statement are discussed in Section~\ref{sec2:probst}. In Section~\ref{sec3:mathback}, we briefly review the mathematical background including max plus algebra. 
Upper and lower bounds on the growth rate for fixed graphs is derived in Section~\ref{sec5:bounds}, and for random graphs in Section~\ref{sec6:randomgraphs}. In Section~\ref{sec:empirical_ub}, we introduce a correction factor on the upper bound. In Section~\ref{sec7:propalg}, we propose a two-run, max-based consensus algorithm robust to additive noise in the network. Simulation results are provided in Section~\ref{sec8:simulations}, followed by conclusions in Section~\ref{sec9:conclusion}.

Vectors are denoted by boldface lower-case, and matrices by boldface upper-case letters. For a matrix $\A$, $[\A]_{i,j}$ denotes the element in the $i^{th}$ row and $j^{th}$ column. 
The symbol $|\cdot|$ denotes absolute value for a real or complex numbers and cardinality for sets. 
Vector \textbf{1} represents a $N\times 1$ column vector of all ones, $[1,1 \dots 1 ]^T$. 
Throughout the paper, $\log(\cdot)$ indicates natural logarithm. We denote the probability density function (PDF) by $f(\cdot)$ and cumulative distribution function (CDF) by $F(\cdot)$.

\section{System Model}
\label{sec2:probst}
We consider a network of $N$ nodes. The communication among nodes is modeled as an undirected graph $\gra=(\ver,\edg)$, where $\ver =\{1,\cdots,N\}$
is the set of nodes and $\edg$ is the set of edges connecting the nodes.  
The set of neighbors of node $i$ is denoted by $ \neig_i = \{ j | \{i,j\} \in \edg \}$. The degree of the $i^{th}$ node, denoted by $d_i=|\neig_i|$, is the number of neighbors of the $i^{th}$ node. The degree matrix $\D$, is a diagonal matrix that contains the degrees of the nodes along its diagonal. The connectivity structure of the graph is characterized by the adjacency matrix $ \A$, with entries $[\A]_{i,j}=1$ if $\{i,j\} \in \edg$ and $[\A]_{i,j}=0$, otherwise.
\textit{Spectral radius} of the network $\rho$, corresponds to the eigenvalue with the largest magnitude of the adjacency matrix \textbf{A}.

In this paper, we consider the following standard assumptions on the system model :
\begin{enumerate}[i)]
	\item Each node has a real number which is its own initial measurement.
	\item At each iteration, nodes broadcast their state values to their neighbors in a synchronized fashion \cite{a2_nejad2009max,aa1_nejad2010max}. Our analysis and the algorithm can be extended to asynchronous networks, assuming that the communication time is small such that the collisions are absent between communicating nodes \cite{a1_iutzeler2012analysis,r1_improved_maxbounds}.
	\item Communications between nodes is analog \cite{intro4_olfati2007consensus,model1_kar2009distributed,a1_iutzeler2012analysis,a2_nejad2009max,sys1_li2010consensus} over the wireless channel and is subject to additive noise. 
	\item General model of time-varying graphs are considered, wherein, a message corrupted by additive noise is received with a probability $1-p$, in order to model the imperfect communication links.
\end{enumerate}
A system model with imperfect transmissions is considered in \cite{r1_improved_maxbounds,aa1_nejad2010max}, where a message is received with a probability $1-p$, unaffected by the communication noise. 
Note that, ours is a more general model that not only consider transmission failures (erasures), but also the errors in transmission due to imperfect communication links or fading channels. The system models used in different applications such as distributed max plus systems \cite{sys3_farahani2017optimization,sys4_fidler2018non}, distributed detection and target tracking \cite{sys5_hu2010distributed}, distributed sensor fusion \cite{sys2_zhu2018mitigating} and multi-agent control systems \cite{sys1_li2010consensus} literature resembles our model.

\subsection{Problem Statement}
Our goal is to have each node reach consensus on the maximum of the node initial measurements in a distributed network, in the presence of additive communication noise. In existing max consensus algorithms \cite{a1_iutzeler2012analysis,r1_improved_maxbounds,max3_shi2012finite,a3_giannini2016asynchronous,a2_nejad2009max,aa1_nejad2010max}, at each iteration a node updates its state value by the maximum of the received values from its neighbors. 
After a number of iterations which is on the order of the diameter of the network, each node reaches a consensus on the maximum of the initial measurements.
However, this approach fails in the presence of additive noise on the communication links, because every time a node updates its state value by taking the maximum over the received noisy measurements, the state value of the node drifts. 

To address this problem, we use max-plus algebra and large deviation theory to find the growth rate of the state values. We then propose an algorithm which locally estimates the growth rate and updates the state values accordingly to reach consensus on the true maximum value.

\section{Mathematical Background}
\label{sec3:mathback}
For completeness, we briefly review the mathematical background
including the max-based consensus algorithm and max-plus algebra.

\subsection{Review of max-based consensus algorithm}
In this section, we describe the conventional max-based consensus algorithm.
Consider a distributed network with $N$ nodes with real-valued initial measurements, $\mathbf{x}(0)=[x_1(0),\dots,x_N(0)]^T$, where $x_i(t)$ denotes the state value of the $i^{th}$ node at time $t$. Max consensus in the absence of noise merely involves updating the state value of nodes with the largest received measurement thus far in each iteration so that the nodes reach consensus on the maximum value of the initial measurements. Let $v_{ij}(t)$ be a zero mean, independent and identically distributed (i.i.d) noise sample from a general noise distribution, which models the additive communication noise between nodes $i$ and $j$ at time $t$. To reach consensus on the maximum of the initial state values, nodes update their state by taking the maximum over the received measurements from neighbors and their own state, given by,
\begin{equation}
\label{eqn:mws_max}
x_i(t+1)={\rm max}\big(x_i(t),\underset{j\in \neig_i}{\rm max}(x_j(t)+v_{ij}(t))\big).
\end{equation}

\subsection{Review of max plus algebra}
We briefly introduce max plus algebra which can be used to represent max consensus algorithm as a discrete linear system. A max-plus approach was considered for max consensus in \cite{a2_nejad2009max,aa1_nejad2010max}, but in the absence of additive noise. Our approach here in considers the presence of a general noise distribution and study its effects on equation~(\ref{eqn:mws_max}) using max-plus algebra and subadditive ergodic theory. 

Max plus algebra is based on two binary operations, $\oplus$ and $\otimes$, on the set $\mathbb{R}_{\rm max}= \mathbb{R} \cup \{-\infty\}$. The operation are defined on $x,y \in \mathbb{R}_{\rm max}$ as follows,
\begin{equation*}
x \oplus y = {\rm max}(x,y)\hspace{1cm}{\rm and} \hspace{1cm}x \otimes y = x +y .
\end{equation*}
 The neutral element for the $\oplus$ operator is $\varepsilon := -\infty$
and for $\otimes$ operator is $e:=0$. Similarly for matrices $\mathbf{X},\mathbf{Y} \in \mathbb{R}_{\rm max}^{N \times N}$, operations are defined as, for $i = 1,\dots,N.$ and $  j=1,\dots,N$.
\begin{align*}
& [\mathbf{X} \oplus \mathbf{Y}]_{i,j}=[\mathbf{X}]_{i,j} \oplus [\mathbf{Y}]_{i,j},  \\
& [\mathbf{X} \otimes \mathbf{Y}]_{i,j} = \bigoplus_{k=1}^N([\mathbf{X}]_{i,k} \otimes [\mathbf{Y}]_{k,j}) = \underset{k}{\rm max} ([\mathbf{X}]_{i,k}+[\mathbf{Y}]_{k,j}), 
\end{align*}
where $[\mathbf{X}]_{i,j}$ and $[\mathbf{Y}]_{i,j}$ denote $(i,j)$ element of matrices $\mathbf{X}$ and $\mathbf{Y}$, respectively.
For integers $k>l$, we denote $\mathbf{Y}(k,l)=\mathbf{Y}(k)\otimes\mathbf{Y}(k-1)\otimes\dots\mathbf{Y}(l)$. 

Consider $\mathbf{x}(t)$ to be an $N \times 1$ vector with the state values of the nodes at time $t$. We can use max plus algebra to represent equation (\ref{eqn:mws_max}) as,
\begin{align}\label{eqn:maxplus rec}
\mathbf{x}(t+1)&=\Snoise(t)\otimes\mathbf{x}(t),\hspace{0.5cm} t>0, \\
&=\underbrace{\Snoise(t)\otimes\Snoise(t-1)\otimes\dots\Snoise(0)}_{\triangleq \Snoise(t,0)} \otimes\mathbf{x}(0), \nonumber
\end{align} 
where $\Snoise(t)$ is the $N \times N$ noise matrix at time $t$, with elements
\begin{equation}
\label{s_mws}
[\Snoise(t)]_{i,j}=
\begin{cases}
e & i=j, \\
\varepsilon, \hspace{0.2cm} & {\rm if} \hspace{0.2cm}\{i,j\}\notin \edg\\
v_{ij}(t), & {\rm if} \hspace{0.2cm}\{i,j\}\in \edg 
\end{cases}
\end{equation}

\subsection{Existence of linear growth}
In a queuing theory and networking context, reference ~\cite{b1_van2007ergodic},\cite{a5_heidergott2006max} show that for a system represented by the recursive relation in equation (\ref{eqn:maxplus rec}), $x_i(t)$ grows linearly, in the sense there exists a real number $\lambda$ such that, for all $i=1,\dots,N$,
\begin{equation}\label{eqn:def_of_lyp}
\lambda = \underset{t \to \infty}{\rm lim}\frac{1}{t}x_i(t), \hspace{0.7cm} {\rm and} \hspace{0.7cm}
\lambda = \underset{t \to \infty}{\rm lim}\frac{1}{t}\mathbb{E}[x_i(t)],
\end{equation} 
where the first limit converges almost surely.
Note that the constant $\lambda$ does not depend on the initial measurement $\mathbf{x}(0)$, or the node index $i$. It is also sometimes referred to as the max-plus \textit{Lyapunov exponent} of the recursion in equation~(\ref{eqn:maxplus rec}).

In our current WSN context, the growth of $x_i(t)$ is clearly dependent on the distribution of noise and graph topology. However, there exists no analytical expressions for the growth rate $\lambda$, even for the simplest graphs and noise distributions. Indeed this is related to a long-standing open problem in the first and last passage percolation \cite{p1_ES_auffinger201550} to obtain analytical expressions for $\lambda$. 
One of our main contributions herein is analytical bounds on $\lambda$ for arbitrary graphs and general noise distributions. We introduce theorems to upper and lower bound the growth rate for arbitrarily connected fixed and random graphs. 

\section{Bounds on growth rate for fixed graphs}
\label{sec5:bounds}

\subsection{Upper bound}
To derive our upper bound on the growth rate, we provide the following theorem for fixed graphs and general noise distributions.
Before stating the theorem, we introduce the following Lemma which will be later invoked in the theorem.
\begin{lemma}
	Let $\A$ be the adjacency matrix and $\rho$ be the spectral radius, then $[\A^t]_{i,j} \le \rho^t$. 
\end{lemma}	
{\rm \textbf{Proof: }}{Consider a singular value decomposition (SVD) of $\A = \mathbf{U}\boldsymbol{\Sigma}\mathbf{V}^T$, so that $\A^t = (\mathbf{U}\boldsymbol{\Sigma}\mathbf{V}^T)(\mathbf{U}\boldsymbol{\Sigma}\mathbf{V}^T)\cdots (\mathbf{U}\boldsymbol{\Sigma}\mathbf{V}^T)$, $t$ times. Let $\mathbf{e}_i$ be a unit vector of zeros, except a $1$ at the $i^{th}$ position. Hence, we can write, $[\A^t]_{i,j} = \mathbf{e}_i^T(\mathbf{U}\boldsymbol{\Sigma}\mathbf{V}^T)^t \mathbf{e}_j$ and show that $[\A^{t}]_{i,j} \le \rho^{t}$ by showing $|\mathbf{e}_i^T \rho^{-t} \A^t \mathbf{e}_j| \le 1$. To this end, we write,
	$$
	\rho^{-t} \A^t = (\mathbf{U}\boldsymbol{\bar{\Sigma}}\mathbf{V}^T)^t,
	$$
	where, $\boldsymbol{\bar{\Sigma}} = \rho^{-1} \boldsymbol{{\Sigma}}$ is a diagonal matrix with diagonal elements $(1,\frac{\rho_2}{\rho},\cdots,\frac{\rho_N}{\rho})$, where $\rho_n$ is the $n^{th}$ largest singular value of $\Sigma$.
	Since $\mathbf{U}$ and $\mathbf{V}^T$ are unitary, it is clear that $\boldsymbol{\bar{\Sigma}}$ is a contraction so that
	\begin{equation}\label{eqn:svd}
	||\mathbf{U}\mathbf{x}||= ||\mathbf{x}||,~~ ||\mathbf{V^T}\mathbf{x}||= ||\mathbf{x}||,~~ ||\mathbf{\boldsymbol{\bar{\Sigma}}}\mathbf{x}|| \le ||\mathbf{x}||,
	\end{equation}
	because $|\frac{\rho_n}{\rho}| < 1$, for $n = 2,\cdots,t$.
	Now, successive application of equation~(\ref{eqn:svd}) yields,
	\begin{align*}
	\mathbf{e}_i^T \rho^{-t} \A^t \mathbf{e}_j &= |\mathbf{e}_i^T \rho^{-t} \A^t \mathbf{e}_j| \\
	& = |\mathbf{e}_i^T (\mathbf{U}\boldsymbol{\bar{\Sigma}}\mathbf{V}^T \cdots \mathbf{U}\boldsymbol{\bar{\Sigma}}\mathbf{V}^T) \mathbf{e}_j| \le 1.
	\end{align*}
	where the first equality is because $\A$ has non-negative entries and the inequality uses equation~(\ref{eqn:svd}) and Cauchy-Schwartz inequality. 
	Hence, $[\A^t]_{i,j} \le \rho^t$, which concludes the proof of Lemma $1$.\\} 
 
\begin{thm}(Upper Bound)
	\label{the1}
	Suppose the moment generating function of the noise $M(\gamma):= \mathbb{E}[e^{\gamma v_{ij}(t)}]$ exists for $\gamma$ in a neighborhood of the origin. Then, 
	an upper bound on growth rate $\lambda$ is given by, 
	\begin{equation}\label{eqn:the_ub}
	\lambda \le {\rm inf}\bigg\{ x : \underset{0\le \beta \le 1}{\rm sup} \bigg[ H(\beta) + \beta \log(\rho) - \beta I\bigg(\frac{x}{\beta}\bigg) < 0\bigg]  \bigg\},
	\end{equation}
	where, $\rho$ is the spectral radius of the graph, $H(\beta)$ is the binary entropy function given by 
	$$
	H(\beta) = -\beta \log(\beta)-(1-\beta)\log(1-\beta),
	$$
	and $I(x)$ is the large deviation rate function of the noise, given by,
	$$
	I(x) := \underset{\gamma > 0}{\rm sup} (x \gamma - \log (M(\gamma))).
	$$  
\end{thm}
{\textbf{Proof:} 
	We  begin by describing the approach taken to prove the theorem.
	We start with formulating growth rate $\lambda$ as a function of the maximum path sum of random variables. Next, to find the maximal path sum, we count the number of paths in $t$ hops that involves $l$ self-loops. We then put the upper bound in the desired form using large deviation theory. The different parts of the proof are labeled accordingly, for readability.
	\\ \textbf{Relate $\lambda$ and maximal path sum :}
To prove Theorem~$1$, we upper bound $\lambda$ using the elements of $\Snoise(t,0)$ defined in equation~(\ref{eqn:maxplus rec}). The $i,j$ entry $[\Snoise(t,0)]_{i,j}$ can be written as the maximum of the sum of noise samples over certain paths. To be precise, let $\pathset_t(i,j)$ be the set of all path sequences $\{\path(k)\}_{k=0}^t$, that start at $p(0)=j$ and end at $p(t)=i$, and also satisfies $\big(\path(k),\path(k+1)\big) \in \edg$ or $\path(k) = \path(k+1)$ for $k \in \{0,1,\cdots,t-1\}$, which allows self loops.
For simplicity we define $\lyap^{(i,j)} \triangleq \big[ \Snoise(t,0)\big]_{i,j}$. The path sum $\lyap^{(i,j)}$ corresponds to the path whose sum of i.i.d noise samples along the edges in $t$ hops between nodes $i$ and $j$, is maximum among all possible paths, is given by,  
	\begin{equation}\label{eqn:def_lyap}
	\lyap^{(i,j)} \triangleq \big[ \Snoise(t,0)\big]_{i,j} = \underset{\{\path(k)\} \in \pathset_t(i,j)}{\rm max} \sum_{k=0}^{t} \big[\Snoise(k)\big]_{\path(k),\path(k+1)}.
	\end{equation}
	For the system defined by the recursive relation in equation~(\ref{eqn:maxplus rec}), let us define the growth rate of this max-plus process to be $\lambda$ and derive an upper bound on $\lambda$. We can relate $\lambda$ to $\lyap^{(i,j)}$ by first recalling the definition in equation~(\ref{eqn:def_of_lyp}),
	\begin{align}\label{eqn:mij_oft}
	 \lambda & =\underset{t \to \infty}{\rm lim}\frac{1}{t}x_i(t)= \underset{t \to \infty}{\rm lim}\frac{1}{t}~\underset{j}{\rm max} \big(\lyap^{(i,j)}+x_j(0)\big).\\ \nonumber
	& \le \underset{j}{\rm max}\bigg(\underset{t \to \infty}{\rm lim~sup} \frac{1}{t} \lyap^{(i,j)} + \underset{t \to \infty}{\rm lim~sup}\frac{x_j(0)}{t}\bigg) \\ \nonumber
	& \le \underset{j}{\rm max} ~\underset{t \to \infty}{\rm lim~sup} \frac{1}{t} \lyap^{(i,j)}. \\ \nonumber
	\end{align}
	In fact, Kingman's subadditive ergodic theorem can be invoked \cite{b1_van2007ergodic} to show that the $\limsup$ in the last inequality be replaced by a limit. Furthermore, as shown in same reference, this limit is independent of $i$, and $j$. Hence, one can work with $\lyap^{(i,j)}$ instead of $x_i(t)$ to upperbound the graph-dependent constant $\lambda$. This enables us to drop the maximum over $j$ and study the constant that $\lyap^{(i,j)}/t$ converges to. Toward this goal, consider the smallest value of $x$ for which
	
	\begin{equation}\label{eqn:lim_ub}
	\underset{t \to \infty}{\rm lim}~P\bigg[\frac{1}{t} \lyap^{(i,j)} > x\bigg] = 0 .
	\end{equation}
	We will upperbound this probability to find bounds on such values of $x$.
	\\
	\textbf{Count the number of paths with $l$ self-loops :}
	Examining equation~(\ref{eqn:def_lyap}) we observe that, for a self-loop at time $k$, $\path(k)=\path(k+1)$. Since $[\Snoise]_{i,i}(k)=e \equiv 0$, there is no contribution to the sum in equation~(\ref{eqn:def_lyap}), as self-loops are not affected by the noise. So it is useful to express the maximum in equation~(\ref{eqn:def_lyap}) over the paths that have a fixed number of self-loops $l$. 	
	To study this case, first we need to count the number of paths that contain $l$ self loops.
	Consider the expression $[(\A+z\mathbf{I})^t]_{i,j}$ where $z$ is an indeterminate variable that will help count the number of paths from node $i$ to node $j$ in $t$ steps that go through a fixed number of $l$ self-loops. 
	Using the binomial expansion we can write,
	\begin{equation*}
	[(\A+z\mathbf{I})^t]_{i,j}=\sum_{l=0}^{t}z^l [\A^{t-l}]_{i,j} {t \choose l}
	\end{equation*}
	where co-efficient of $z^l$ is the number of paths from node $i$ to $j$ in $t$ steps, that go through $l$ self loops denoted as $n_l = {t \choose l} [\A^{t-l}]_{i,j}$. \\
	\textbf{Upper bound the growth rate $\lambda$ : }
	Now we can write,
	\begin{equation}\label{eqn:mij_exp}
	\frac{1}{t} \lyap^{(i,j)} = \underset{l \in \{0,1,\cdots,t-1\}}{\rm max}~{\rm max}\bigg(\frac{S_1^{(l)}}{t},\cdots,\frac{S_{n_l}^{(l)}}{t}\bigg)
	\end{equation}
	where $S_q^{(l)}$ is any sum in equation~(\ref{eqn:def_lyap}) that involves $l$ self loops, $q \in \{1,\cdots,n_l\}$ and $n_l$ is the number of paths in $\pathset_t(i,j)$ with $l$ self-loops. Substituting equation~(\ref{eqn:mij_exp}) into equation~(\ref{eqn:lim_ub}) and using the union bound, we can upper bound equation~(\ref{eqn:lim_ub}) as, 	
	
	\begin{eqnarray}
	\label{eqn: rewrite_union bound}
	P\bigg[\underset{l}{\rm max}\hspace{0.2cm}{\rm max} \bigg( \frac{S_1^{(l)}}{t},\dots,\frac{S_{n_l}^{(l)}}{t}\bigg) > x \bigg] \le \nonumber\\	
	\sum_{l=0}^{t} \sum_{q=1}^{n_l} P\bigg[\frac{1}{t} S_q^{(l)} > x\bigg].
	\end{eqnarray} 
	Since $S_q^{(l)}$ are sum of $(t-l)$ i.i.d random variables, $S_q^{(l)}$ is i.i.d in $q$ for a fixed $l$, but differently distributed for different $l$, so we can drop the index $q$ and replace the sum over $q$ with $n_l$ to get, 
	\begin{align}
	\label{simp_uni_bound}
	P \bigg[\frac{1}{t} \lyap^{(i,j)} > x\bigg] &\le \sum_{l=0}^{t} n_l \cdot P\bigg[\frac{1}{t} S^{(l)} > x\bigg], \nonumber\\
	& = \sum_{l=0}^{t} {t \choose l} [\A^{t-l}]_{i,j}~P\bigg[\frac{1}{t} S^{(l)} > x\bigg].
	\end{align}
	From Lemma~$1$, $[\A^{t-l}]_{i,j} \le \rho^{t-l}$ and letting 
	\begin{equation*}
	l^* = \underset{l}{\rm argmax} {t \choose l} \rho^{t-l} ~P\bigg[\frac{S^{(l)}}{t} > x\bigg]
	\end{equation*}
	in equation (\ref{simp_uni_bound}) we have,
	\begin{equation}
	\label{eqn: reduc_lstar}
	P \bigg[\frac{1}{t} \lyap^{(i,j)} > x\bigg] \le (t+1)~{t \choose l^*} \rho^{t-l^*} P\bigg[\frac{S^{(l^*)}}{t} > x\bigg]
	\end{equation}
	We can rewrite $P\bigg[\frac{S^{(l^*)}}{t} > x\bigg]$ as $P\bigg[\frac{S^{(l^*)}}{t-l^*} > \frac{t}{t-l^*}x\bigg]$. In the next step, we bound the second term on RHS of equation (\ref{eqn: reduc_lstar}) by the Chernoff bound as,
	\begin{equation*}
	P\bigg[\frac{S^{(l^*)}}{t-l^*} > \frac{t}{t-l^*}x\bigg] = e^{-t(1-\alpha)I(\frac{x}{1-\alpha})} 
	\end{equation*}
	where $I(x)$ is the large deviation rate function and $\alpha = l^*/t$. From \cite[pp 666]{pf2_cover2012elements}, for large $t$, we have ${t \choose \alpha t}= e^{(t(H(\alpha)+o(1))}$, where  $H(\alpha) = -\alpha \log(\alpha)-(1-\alpha)\log(1-\alpha)$. For convenience let $\beta = 1-\alpha$, then equation (\ref{eqn: reduc_lstar}) reduces to,
	\begin{equation}
	\label{eqn:reduced_eqn_exp}
	P \bigg[\frac{1}{t} \lyap^{(i,j)} > x\bigg] \le (t+1) e^{t \big( H(\beta) + \beta \log(\rho) - \beta I(x/\beta)+o(1)\big)}
	\end{equation}	
	It is well-known that the large-deviation rate function $I(\cdot)$ is monotonically increasing to infinity for arguments restricted above the mean of the random variable (zero-mean noise in our case) \cite{der1_lewis1997introduction}, so the exponent in equation~(\ref{eqn:reduced_eqn_exp}) will be negative when $x$ is large enough. Hence the smallest $x$ for which equation~(\ref{eqn:reduced_eqn_exp}) goes to zero exponentially is given by equation~(\ref{eqn:the_ub}).
	This concludes the proof of the Theorem. \\}

\subsubsection{Simplified upper bound for Gaussian noise}\label{subsec:ub_Gaus}
If the noise is Gaussian, i.e $v_{ij} \sim \mathcal{N}(0,1)$, then $I(x) = \frac{x^2}{2}$ in equation~(\ref{eqn:the_ub}). Using algebra, equation~(\ref{eqn:the_ub}) simplifies as, 
\begin{eqnarray}\label{eqn:ub_gauss}
\lambda \le \underset{0 \le \beta \le 1}{\rm sup} \sqrt{2 \beta \big(H(\beta)+ \beta \log(\rho) \big)}.
\end{eqnarray}
Defining $g(\beta) = \sqrt{2 \beta \big(H(\beta)+ \beta \log(\rho) \big)}$,  the supremum will be achieved for $\beta$ that satisfies $\frac{\partial g(\beta)}{\partial \beta}=0$,
which simplifies to
$$
\rho = \sqrt{\frac{\beta}{1-\beta}} e^{-\frac{H(\beta)}{2\beta}}.
$$
Note that, $I(\cdot)$ is a convex function and as $\rho$ increases $\beta$ will approach its upper limit of $1$. Therefore, we can conclude that for graphs with large $\rho$, the optimal value of $\beta \to 1$, hence we can write,
\begin{equation}\label{eqn:gauss_ub_approx}
H(\beta) + \beta \log(\rho) - \beta I(x/\beta) \approx \log(\rho) - I(x)	
\end{equation}
which is negative when $I(x) > \log(\rho) $. 

We established this behavior of $\beta$ for the Gaussian case. However this holds more generally. Since $f(x,\beta) = H(\beta) + \beta \log(\rho) - \beta I(x/\beta)$ is concave in $\beta$ for every $x$, we only need to check when $x>0$, the $\beta^*$ that solves $\frac{\partial f(x,\beta^*)}{\partial \beta}=0$ approaches $1$ as $\log(\rho)$ increases. Setting the derivative to $0$, we get, 
	$$
	\log\bigg(\frac{1-\beta}{\beta}\bigg)+\log(\rho)-I(x/\beta)+\frac{x}{\beta} I^{'} (x/\beta) = 0.
	$$
	One can check that as $\rho$ increases, $\log(\rho) \to \infty$ and hence, we need $\log\big(\frac{1-\beta}{\beta}\big)\to -\infty$ which is reached as $\beta \to 1$.
	 This shows that as $\rho$ increases, $\beta \to 1$ for general noise distributions as well. \\
\subsubsection{Alternative upper bound}
Recall that, while proving Theorem~\ref{the1}, we were interested in the path from node $i$ to $j$ in $t$ steps, whose sum was the maximum among all possible paths. To achieve this, first we had to count the number of paths from node $i$ to $j$ in $t$ steps and then, group these paths in terms of number of paths that involved self-loops. Note that, self-loops were not affected by noise so their contribution to the sum along the path is $0$. The analysis would be simpler if we considered noise on self loops, thereby eliminating the need to count and group the paths by number of self loops involved. So considering noise on self-loops, which is equivalent to setting $\beta = 1$ in Theorem~\ref{the1}, would result in the following recursion,
\begin{equation}
\label{eqn:alt_ubrec}
x_i(t+1)={\rm max}\big(x_i(t)+v_{ii}(t),\underset{j\in \neig_i}{\rm max}(x_j(t)+v_{ij}(t))\big)
\end{equation}
instead of equation (\ref{eqn:mws_max}). Note that, equation (\ref{eqn:alt_ubrec}) is not the proposed max consensus scheme, but an auxiliary recursion used here to upper bound the growth rate. We can observe that $x_i(t+1)$ is convex in $v_{ii}(t)$, and due to Jensen's inequality the additional noise in equation (\ref{eqn:alt_ubrec}) can only increase the slope $\lambda$ compared to equation (\ref{eqn:mws_max}). Hence, the growth rate of equation~(\ref{eqn:mws_max}) is upper bounded by that of equation~(\ref{eqn:alt_ubrec}). Repeating the proof of Theorem~\ref{the1} for this case amounts to replacing $\A$ by $\A + \mathbf{I}$ and therefore $\rho$ with $\rho + 1$, so we have the following :

\begin{thm}\label{the2:altub}
	The auxiliary recursion in equation (\ref{eqn:alt_ubrec}) has a growth rate upper bounded by the value of $x > 0$ that solves,
	\begin{equation}\label{eqn:aub_conv}
		I(x)=\log(\rho+1),
	\end{equation}
	where $I(x)$ is the large deviation rate function. Moreover, this value of $x$ upper bounds the growth rate $\lambda$ of the recursion in equation (\ref{eqn:mws_max}).
\end{thm}

Note that, for Gaussian noise distribution the alternative upper bound on the growth rate can be calculated as,
\begin{equation}\label{eqn:alt_up_g}
\lambda \le \sqrt{2 \log(\rho + 1)}.
\end{equation} 
While equation~(\ref{eqn:alt_up_g}) is a looser bound than equation~(\ref{eqn:ub_gauss}), it is much simpler. 
We find that as $\rho$ increases, i.e as $\beta \to 1$, alternative upper bound and exact upper bound converge. 

\subsection{Lower bound}
While it is clear that $\lambda \ge 0$, it is not obvious when $\lambda > 0$. In this section, we derive lower bound, which, in part, shows that there exists a growth rate $\lambda$ due to additive noise in the network, which is always positive $(\lambda > 0)$. Also, the lower bound relates to the order statistics of the underlying noise distribution as well as the steady state distribution of the underlying Markov chain.\\ 

\subsubsection{Lower bound for regular graphs}
Recall that the state of the $i^{th}$ sensor at time $t+1$ is given by the $i^{th}$ element of the vector,
$
\mathbf{x}(t+1)=\Snoise(t,0) \otimes \mathbf{x}(0)
$
which is,
\begin{align}
x_i(t+1)&=\underset{j}{\rm max}\big(\big[\Snoise(t,0)\big]_{i,j}+x_j(0)\big), \nonumber \\
&\ge \underset{j}{\rm max}\big[\Snoise(t,0)\big]_{i,j}+x_{\rm min}(0), \label{eqn:lb_ineq}
\end{align}
where $x_{\rm min}(0)=\underset{i}{\rm min}~x_i(0)$.
Now, using equation~(\ref{eqn:lb_ineq}), we can lower bound the growth rate $\lambda$ as,
\begin{subequations}
	\begin{align}
	\lambda &= \underset{t \to \infty}{\rm lim}\frac{x_i(t)}{t} = \underset{t \to \infty}{\rm lim}\frac{x_i(t+1)}{t} \nonumber \\
	& \ge \underset{t \to \infty}{\rm lim}\frac{1}{t}~\underset{j}{\rm max}\big[\Snoise(t,0)\big]_{i,j}+  \underset{t \to \infty}{\rm lim}\frac{1}{t}~x_{\rm min}(0) \label{eqn:subeq1}\\
	& \ge \underset{t \to \infty}{\rm lim}\frac{1}{t}~\sum_{k=0}^{t-1}\big[\Snoise(k)\big]_{\path(k),\path(k+1)} \label{eqn:lb_subeq2}.
	\end{align} 
\end{subequations}
where equation~(\ref{eqn:subeq1}) is due to equation~(\ref{eqn:lb_ineq}) and in equation~(\ref{eqn:lb_subeq2}), $\{\path(k)\}_{k=0}^t$ is \textit{any} path that satisfies $\path(0)=j$ and $\path(t)=i$. In order to get a good lower bound, we rely on evaluating equation~(\ref{eqn:lb_subeq2}) for a specific path defined as,
\begin{equation}
\label{eqn:path1}
\path(k+1)=\underset{m ~\in~ \neig(\path(k))~{\rm \cup}~ \path(k)}{\rm argmax} \big[\Snoise(k)\big]_{\path(k),m}.
\end{equation}
This amounts to selecting the locally optimum or greedy path. If the graph is $d-$regular, then with $\path(k)$ chosen as in equation~(\ref{eqn:path1}), the random variables in equation~(\ref{eqn:lb_subeq2}) are distributed the same as the maximum of $d$ i.i.d random variables and zero, whose expectation is denoted as $\exmax(d)$. Therefore, due to law of large numbers, equation~(\ref{eqn:lb_subeq2}) converges to,
\begin{align}
\lambda \ge \exmax(d) &= \mathbb{E} \bigg[{\rm max} \bigg(0, \underset{m}{\rm max}\big[\Snoise(k)\big]_{\path(k),m} \bigg)\bigg],\nonumber\\
& = d~\int_{0}^{\infty} x~F^{d-1}(x)f(x)~dx.
\end{align}
where $F(\cdot)$ and $f(\cdot)$ are the CDF and PDF of the noise respectively. Also, using \cite[pp 80]{lb1_david2004order}, one can lower bound growth rate with a simpler expression given by,
$$
\lambda \ge F^{-1}\bigg(\frac{d}{d+1}\bigg),
$$
provided that median of noise samples are zero. \\

\subsubsection{Lower bound for irregular graphs}
For irregular graphs, the path defined in equation~(\ref{eqn:path1}) is a random walk on the graph with the corresponding sequence of nodes constituting a Markov chain. When the graph is irregular, the transition probabilities of this Markov chain depend on the degree of the current node. Specifically, the transition probability matrix is given by,
$$
\mathbf{P} = (1-\kappa)\D^{-1}\A+\kappa~\mathbf{I}
$$ 
where the diagonal matrix $[\D]_{i,i}=d_i$, degree of node $i$, so that 
\begin{equation}\label{eqn:tr_prob}
[\mathbf{P}]_{i,j}  = 
\begin{cases}
\frac{1-\kappa}{d_i} & i \neq j, \hspace{0.2cm}(i,j) \in \edg \\
\kappa & i=j, \\
\end{cases}
\end{equation}
where $\kappa$ is the probability that noise samples on neighboring edges of node $i$ are negative, given by
\begin{eqnarray*}
	\kappa = P\bigg[[\Snoise(k)]_{i,j}<0, \forall j \bigg] = d_i \int_{-\infty}^{0} F^{d_i-1}(x)~ f(x)~dx.
\end{eqnarray*}
Let the steady state probabilities of this Markov chain be denoted by $\pi_i$. Then, using the law of large numbers the lower bound is given by,
\begin{equation}\label{eqn:low_pionly}
\underset{t \to \infty}{\rm lim} \frac{1}{t} \sum_{k=0}^{t-1} \underset{m}{\rm max}\big( [\Snoise(k)]_{\path(k),m}\big) = \sum_{i=1}^{N} \pi_i~\exmax(d_i),
\end{equation} 
since the random variable $\underset{m}{\rm max}\big( [\Snoise]_{\path(k),m}\big)$ has expectation $\exmax(d_i)$, given node $i$. One can find a closed form expression for $\pi_i$ as $\pi_i = \frac{d_i}{2E}$ \cite[pp 78]{pf2_cover2012elements}, where $E:=|\edg|$ is the total number of edges in the network.  
To verify this, one can check that $\boldsymbol{\pi}^T\mathbf{P} = \boldsymbol{\pi}^T$, where $\boldsymbol{\pi}^T = [{\pi}_1,\cdots,\mathbf{\pi}_N]$, using equation~(\ref{eqn:tr_prob}). In conclusion, the lower bound on the growth rate for irregular graphs is given by,
\begin{equation}\label{eqn:lowb_irreg}
\lambda \ge \sum_{i=1}^{N} \frac{d_i}{2E} \exmax(d_i).
\end{equation}

\section{Bounds on growth rate for Random Graphs}
\label{sec6:randomgraphs}
In this section we consider the case where each edge is absent by a probability of $p$, independently across edges and time, which models random transmission failures.  

\subsection{Upper bound for random graphs}
We now show that the upper bound on growth rate for the randomly changing graphs can be simply obtained by replacing $\rho$ in the fixed graph case by $\rho(1-p)$ in equation (\ref{eqn:the_ub}), where $p$ is the Bernoulli probability, that any edge will be deleted independently at each iteration. 

Recall that in fixed graph model, $\Snoise(k)$ had zero $(e)$ along the diagonal and $[\Snoise(k)]_{l,m} = v_{lm}(k)$ was the underlying i.i.d noise random variables when $(l,m)~\in~\edg$. The random graph can be described as,
\begin{equation}
\label{eqn:def_random}
[\Snoise(k)]_{l,m}=
\begin{cases}
v_{lm}(k) & {\rm with~prob~}{(1-p)}{\rm~if~}(l,m)~\in~\edg \\
-C & {\rm with~prob~}{p}{\rm~if~}(l,m)~\in~\edg \\
e & l=m, \\
\varepsilon & {\rm if~}(l,m)\notin \edg\\
\end{cases}
\end{equation}
where $C$ is a large positive constant which captures randomly absent edge as $C \to \infty$. Note that, since each node is maxing with itself at each iteration in equation~(\ref{eqn:mws_max}), the large negative value of $-C$, will never propagate through the network, which is equivalent to deleting an edge, for large $C$.  

Following the analysis of the fixed graph case, only the moment generating function of the noise samples changes to,
\begin{equation*}
M(\gamma,C) = p e^{-C\gamma} + (1-p) M(\gamma),
\end{equation*}
where $M(\gamma)$ is the original moment generating function of the noise samples given by $M(\gamma) = \mathbb{E}\big[e^{\gamma v_{ij}(k)} \big]$. The corresponding rate function is given by 
$$
 I(x,C) = \underset{\gamma > 0}{\rm sup}~ \big(x \gamma - \log (M(\gamma,C))\big).
 $$

Following the proof of Theorem~\ref{the1}, to upper bound the growth for this case we have to find the smallest $x$ that satisfies,
$$
\underset{C \to \infty}{\rm lim}~\underset{0 \le \beta \le 1}{\rm sup} \bigg(H(\beta) + \beta\log(\rho)-\beta I\bigg(\frac{x}{\beta},C\bigg) <0 \bigg)
$$
Consider $f(x,\beta,C)=H(\beta) + \beta\log(\rho)-\beta I\big(\frac{x}{\beta},C\big)$, since $f(x,\beta,C)$ is convex in $C$ and concave in $\beta$ we can write,
\begin{align}
\label{eqn:sup_rand}
\underset{C }{\rm inf} ~\underset{0 \le \beta \le 1}{\rm sup} f(x,\beta,C)& = \underset{0 \le \beta \le 1}{\rm sup}~\underset{C }{\rm inf}~f(x,\beta,C). \\
&= \underset{0 \le \beta \le 1}{\rm sup}~\underset{C \to \infty}{\rm lim}f(x,\beta,C). \nonumber\\
=  \underset{0 \le \beta \le 1}{\rm sup}\bigg(H(\beta) &+ \beta\log(\rho(1-p)) -\beta I\bigg(\frac{x}{\beta}\bigg) < 0\bigg), \nonumber
\end{align}
where the first equality is due to classical minimax theorem, and second due to the monotonicity of $f(x,\beta,C)$ in $C$. 
Hence, the upper bound can be written as,
\begin{equation}\label{eqn:ub_randg}
\lambda \le {\rm inf}\bigg\{ x : \underset{0\le \beta \le 1}{\rm sup} \bigg[ H(\beta) + \beta \log(\rho(1-p)) - \beta I\bigg(\frac{x}{\beta}\bigg)<0\bigg]  \bigg\}.
\end{equation}

Interestingly, this is precisely the upper bound for fixed graphs except that we have $\rho(1-p)$ instead of $\rho$. While for a fixed graph $\rho \ge 1$ always holds, in random graphs case it is possible to have $\rho (1-p) < 1$. If $\rho(1-p)\approx 0$ then it is easy to check in equation~(\ref{eqn:ub_randg}) that the optimizing $\beta$ is near zero. This can be contrasted with the case where $\rho$ is large and the optimizing $\beta$ was found to satisfy $\beta \approx 1$ in Section~\ref{subsec:ub_Gaus}.

\subsection{Lower bound for random graphs}
Here, we derive the lower bound on the growth rate for randomly changing graphs. Recall that, for the path defined in equation~(\ref{eqn:path1}), and when $\Snoise(k)$ is as defined in equation~(\ref{eqn:def_random}), yields a lower bound on the growth rate, for graphs with edge deletion probability of $p$.  

Compared to equation~(\ref{eqn:lowb_irreg}), the only difference in the derivation is that, the node $i$ will now have a random degree $Z_i$, which is binomial with parameters $(d_i,1-p)$.
Due to law of large numbers, equations~(\ref{eqn:low_pionly})-(\ref{eqn:lowb_irreg}) have an additional expectation with respect to this binomial distribution, resulting in following expression,
\begin{align}\label{eqn:lowb_rand}
	\lambda &\ge \sum_{i=1}^{N} \pi_i \mathbb{E} \big[\exmax(Z_i)\big] \\
&	= \sum_{i=1}^{N}\frac{d_i}{2E} \sum_{k=0}^{d_i} {d_i \choose k} p^{d_i - k} (1-p)^{k} \exmax(k). \nonumber
\end{align}
Note that, in equation~(\ref{eqn:lowb_rand}), $\pi_i = d_i / 2E$ still holds, since the transition probabilities of the Markov chain are still of the form as in equation~(\ref{eqn:tr_prob}).

\subsection{Upper bound on growth rate without calculating $I(x)$}
In this section, we present a technique to directly calculate the upper bound on growth rate using the moment generating function, without having to compute the large deviation rate function of the additive noise distribution. 

Recall that the upper bound on growth rate is given by equation~(\ref{eqn:ub_randg})
where, $p=0$ for fixed graphs. For convenience, let $K \triangleq \rho(1-p)$ and $\bar{f}(\beta,x) = H(\beta) + \beta \log(K) - \beta I(x/\beta)$. Since, $I(x) = \underset{\gamma > 0}{\rm sup}\big( x\gamma-\log M(\gamma)\big)$, we can write,
\begin{align}\label{eqn:minimax_1}
\underset{0\le \beta\le 1}{\rm sup} \bar{f}(\beta,x)  = \underset{\gamma > 0}{\rm inf}~ \underset{0\le \beta\le 1}{\rm sup} & \big(H(\beta) + \beta \log(K) \\ \nonumber
&- x\gamma + \beta \log M(\gamma) \big).
\end{align}
where, we used minimax theorem to interchange the infimum and supremum, since $\log M(\gamma)$ is always convex. The inner supremum can be solved in closed form as,
\begin{equation}
	\beta^* = \frac{K M(\gamma)}{1+K M(\gamma)},
\end{equation}
which yields,
$$
\underset{0\le \beta\le 1}{\rm sup} \bar{f}(\beta,x)  = \underset{\gamma > 0}{\rm inf}~ \big(H(\beta^*) + \beta^* \log(KM(\gamma)) - x\gamma \big).
$$ 
So we have, 
\begin{align}
	\underset{x}{\rm inf} \bigg\{ x : \underset{0\le \beta\le 1}{\rm sup} \bar{f}(\beta,x) < 0\bigg\} =  \underset{\gamma > 0}{\rm inf}~ &\bigg( \frac{1}{\gamma}H(\beta^*) \\
	&+ \frac{\beta^*}{\gamma} \log( KM(\gamma)) \bigg) \nonumber
\end{align}
Note that $\beta^*$ is also a function of $\gamma$. This technique is very useful to calculate growth rate, when $I(x)$ is difficult to evaluate, or unavailable.

\section{Empirical upper bound on growth rate} 
\label{sec:empirical_ub}
In this Section, we propose an empirical correction factor to the upper bound which improves the tightness of the bound, for all network settings and noise distributions.
In order to improve the tightness of the upper bound, we introduce a correction factor $\phi$ to our upper bound in equation~(\ref{eqn:the_ub}). The correction factor $\phi$ depends only on number of nodes $N$ in the network, given by,
	\begin{equation}\label{eqn:phi}
\phi = 1 - \frac{1}{2\sqrt{N}},
\end{equation}
and multiplies the upper bound in equation~(\ref{eqn:the_ub}).

While we have no proof that this correction will always yield an upper bound,
the choice of $\phi$ was empirically validated over different graph topologies and noise distributions, and in all settings, $\phi$ improved the tightness of the bound. 
Our intuition is that the approximations made in deriving the upper bound leads to a minor deviation in the tightness for smaller $N$, which can be fixed by $\phi$.
Note that, as $N \to \infty$, the compensation variable $\phi \to 1$, hence $\phi  $ mainly contributes for graphs with smaller number of nodes. 

In Section~\ref{sec8:simulations}, we compare the tightness of upper bound in equation~(\ref{eqn:the_ub}) and the empirical bound, illustrating the accuracy of the correction factor $\phi$. 

\begin{algorithm}[]
	\caption{Robust Max consensus Algorithm}\label{algo: max1}
	\begin{algorithmic}[1] 
		\State \textbf{First run ::}
		\State \;\;\;\;\;\textbf{Input}: iterations $=t$, $\#$ of nodes $=N$
		\State \;\;\;\;\;\textbf{Initialization} 
		\State \;\;\;\;\;\; Initialize all nodes to zero, $x_i(0)=0$  
		\State \;\;\;\;\;\textbf{repeat until : $t_{\rm max}$ iterations}
		\State \;\;\;\;\; \textbf{for \{$i=1:N$\}}
		\State \;\;\;\;\; ${x}_i(t)={\rm max}\big(x_i(t),\underset{j\in \neig_i}{\rm max}({x}_j (t-1) + v_{ij}(t-1))\big) $ \label{alg:max_est_firstrun}
		\State \;\;\;\;\; \textbf{end : for}
		\State \;\;\;\;\;\textbf{end : repeat}
		\State \;\;\;\;\;\textbf{growth rate estimate :} $\hat{\lambda}_i(t_{\rm max}) = \frac{{x}_i(t_{\rm max})}{t_{\rm max}}$ \label{alg:rate est}
		\State \textbf{Second run ::}
		\State \;\;\;\;\;\textbf{Input}: $\#$ of nodes $=N$, Initial state : $x_i(0)$  
		\State \;\;\;\;\;\textbf{repeat until : convergence}
		\State \;\;\;\; \textbf{for \{$i=1:N$\}}
		\State \;\; {\small $x_i(t)={\rm max}\big(x_i(t),\underset{j\in \neig_i}{\rm max}({x}_j (t-1) + v_{ij}(t-1))\big)-\hat{\lambda}_i(t_{\rm max}) $ \label{alg:sec run est} }
		\State \;\;\;\;\; \textbf{end : for}
		\State \;\;\;\;\;\textbf{end : repeat}	
	\end{algorithmic}
\end{algorithm}

\section{Robust Max consensus Algorithm}
\label{sec7:propalg}
Max consensus algorithms in existing works \cite{a1_iutzeler2012analysis,a2_nejad2009max,a3_giannini2016asynchronous,max3_shi2012finite,r1_improved_maxbounds} fail to converge in the presence of noise, as there is no compensation for the positive drift induced by the noise. Authors in \cite{a4_zhang2016max} develop a soft-max based average consensus (SMA) approach to approximate the maximum and compensate for the additive noise. However, their algorithm is sensitive to a design parameter, which controls the trade off between estimation error and convergence speed. So, we develop a fast max-based consensus algorithm in this section, which is informed by the fact that there is a constant slope $\lambda$, analyzed in the previous sections, which can be estimated and removed. This makes the algorithm robust to the additive noise in the network. 

If the knowledge of the spectral radius of the network and noise variance is known, then by using Theorem~\ref{the1}, one can closely estimate the growth rate and subtract this value at each node after the node update.
However, the noise variance and the spectral radius are not always known locally at each node. Hence, we propose a fast max consensus algorithm generalized to unknown noise distributions, as described in Algorithm \ref{algo: max1}, where slope is being locally estimated at each node. We also analyze the variance of this estimator in Section~\ref{subsec:var_ana}.

Our algorithm consists of two runs, where in the first run, we initialize the state values of all the nodes to zero and run the max consensus algorithm in the additive noise setting. This can be performed by a simple reset operation, which is available at every node and then initiate the conventional max consensus algorithm. Note that, in this case the true maximum is zero, but due to the additive noise, the state values grow at the rate of $\lambda$. The growth rate estimate for node $i$ is denoted by $\hat{{\lambda}}_i$, is computed locally over $t_{\rm max}$ iterations as,
\begin{equation}\label{eqn:estimator_of_ub}
	\hat{{\lambda_i}}(t_{\rm max})=\frac{1}{t_{\rm max}}x_i(t_{\rm max}),
\end{equation}
the average increment in the state value of node $i$. Note that, this estimation is done locally at every node. Also, the algorithm is memory-efficient, since the history of state values is not used, and only the information of the iteration index and the current state value is needed to estimate the growth rate. 

In the second run, max consensus algorithm is run on the actual measurements to find the maximum of the initial readings. The growth rate estimate $\hat{\lambda}_i$ is used to compensate for the error induced by the additive noise as given in line~(\ref{alg:sec run est}) of Algorithm~$1$.
Note that, the estimator is independent of the type of additive noise distribution. 
 
 \subsection{Performance Analysis}\label{subsec:var_ana}
To address the accuracy of the estimate in equation~(\ref{eqn:estimator_of_ub}) over a finite number of iterations, we use Efron-Stein's inequality \cite{p1_ES_auffinger201550,p3_ES_sridharan2002gentle} to show that the variance of the growth rate estimator $\hat{{\lambda_i}}(t_{\rm max})$ decreases as $\mathcal{O}(t_{\rm max}^{-1})$, where $t_{\rm max}$ is number of hops. For completeness, the Efron-Stein inequality is introduced in the following theorem.

\begin{thm}\label{the3:effstein}
Let $X_1,X_2,\cdots,X_n$ be independent random variables and let $X_q^{'}$ be an independent copy of $X_q$, for $q \ge 1$. Let $Z = f(X_1,X_2,\cdots,X_{q},\cdots,X_n)$ and
$$ 
Z_q^{'} = f(X_1,X_2,\cdots,X_{q-1},X_{q}^{'},X_{q+1},\cdots,X_n), 
$$
then
$$
{\rm Var}(Z) \le \sum_{q=1}^{n} \mathbb{E}\big[\big((Z-Z_q^{'})_{+}\big)^2\big],
$$
where $(Z-Z_q^{'})_{+}={\rm max}(0,Z-Z_q^{'})$.
\end{thm}
{\textbf{Proof:}} Provided in Theorem~$7$ of \cite{p3_ES_sridharan2002gentle}.\\

The following theorem bounds the variance of the growth rate estimator.
\begin{thm}\label{the4:var}
	The Variance of the growth rate estimator $\hat{{\lambda_i}}(t_{\rm max})$ satisfies, 
	$$
	{\rm Var}(\hat{{\lambda_i}}(t_{\rm max})) \le \frac{\sigma^2}{t_{\rm max}},
	$$
	where $t_{\rm max}$ is number of iterations and $\sigma^2 = {\rm Var}(v_{ij}(t))$.
\end{thm} 
 {\textbf{Proof:}}
 Using equation~(\ref{eqn:estimator_of_ub}), and recalling from Theorem~\ref{the1} the expression for $x_i(t_{\rm max})$ with zero initial conditions $x_i(0)=0$ we have
 \begin{align}\label{eqn:gr_est_var}
 \hat{{\lambda_i}}(t_{\rm \tiny max}) &=\frac{1}{t_{\rm max}}\bigg( \underset{\{\path(k)\} \in \underset{j}{\cup} \pathset_{t_{\rm max}}(i,j)}{\rm max} \sum_{k=0}^{t_{\rm max}} \big[\Snoise(k)\big]_{\path(k),\path(k+1)}\bigg).
 \end{align}
 Next, we use Theorem~\ref{the3:effstein} to bound the variance of equation~(\ref{eqn:gr_est_var}). For simplicity of notation, set $Z = \hat{{\lambda_i}}(t_{\rm max})$, which depends on noise samples $v_{ij}(t)$ through $\Snoise(k)$ in equation~(\ref{eqn:gr_est_var}). So the independent random variables $\mathcal{X}=\{X_1,X_2,\cdots,X_n\}$ in Theorem~\ref{the3:effstein} correspond to re-indexing of $v_{ij}(t)$, with $n$ denoting the total number of noise samples that influence $\hat{{\lambda_i}}(t_{\rm max})$, which is approximately $n \approx (t_{\rm max}+1)E$, where $E=|\edg|$ is the total number of edges (the exact value of $n$ depends on the graph topology). We set $Z_q$ to be given by equation~(\ref{eqn:gr_est_var}) when the noise sample $v_{ij}(t)$ corresponding to $X_q$ is replaced by an independent copy $X_q^{'}$. Note that the path that maximizes equation~(\ref{eqn:gr_est_var}) corresponds to a subset $\mathcal{M}(\mathcal{X})$ of $\{1,\cdots,n\}$, with $t_{\rm max}$ elements.
 
If $q \notin \mathcal{M}(\mathcal{X})$, then $(Z-Z_q^{'})_{+}=0$. This is because, if $X^{'}_q \leq X_q$ then the maximal path is unchanged implying $Z-Z_q^{'}=0$. If $X^{'}_q > X_q$ then either the maximal path is unchanged in which case $Z-Z_q^{'}=0$, or  $X^{'}_q$ is sufficiently larger than  $X_q$ to cause the maximal path to change which implies  $Z-Z_q^{'} \le 0$ and therefore $(Z-Z_q^{'})_{+}=0$.
Hence, we consider only $q \in \mathcal{M}(\mathcal{X})$, so that Theorem~\ref{the3:effstein} can be simplified to involve only $t_{\rm max}$ terms rather than $n$ terms:
 \begin{align}\label{eqn:ES_ineq}
 	& {\rm Var}(Z) \le \mathbb{E}_\mathcal{X} \bigg[\sum_{q \in \mathcal{M}(\mathcal{X})} \mathbb{E}\big[\big((Z-Z_q^{'})_{+}\big)^2\big | \mathcal{X}] \bigg],\\
    & = \mathbb{E}_\mathcal{X} \bigg[ \sum_{q \in \mathcal{M}(\mathcal{X})} \mathbb{E}\big[\big((Z-Z_q^{'})_{+}\big)^2 | (X_q \ge X_q^{'}) | \mathcal{X}\big] P[(X_q \ge X_q^{'}) | \mathcal{X}] \nonumber\\
 	& + \sum_{q \in \mathcal{M}(\mathcal{X})} \mathbb{E}\big[\big((Z-Z_q^{'})_{+}\big)^2 | (X_q < X_q^{'}) | \mathcal{X}\big] P[(X_q < X_q^{'}) | \mathcal{X}] \bigg], \nonumber
 \end{align}  
 where the equality is due to the total expectation theorem. 
  Note that, for $q \in \mathcal{M}(\mathcal{X})$ and $X_q < X_q^{'}$, the maximal path remains the same and $(Z-Z_q^{'})_{+}=0$. Using $P[(X_q \ge X_q^{'}) | \mathcal{X}] = 1/2$ in equation~(\ref{eqn:ES_ineq}) reduces to,
 \begin{align}\label{eqn:final_var_eq}
 	&{\rm Var}(Z) 
 	 \le \frac{1}{2}\mathbb{E}_\mathcal{X} \sum_{q \in \mathcal{M}(\mathcal{X})} \mathbb{E}\big[\big((Z-Z_q^{'})_{+}\big)^2|(X_q \ge X_q^{'}) | \mathcal{X} \big] \nonumber	\\
 	 &\le \frac{1}{2} \mathbb{E}_\mathcal{X} \sum_{q \in \mathcal{M}(\mathcal{X})} \mathbb{E}\bigg[\bigg(\frac{1}{t_{\rm max}}(X_q-X_q^{'})_{+}\bigg)^2 |(X_q \ge X_q^{'}) | \mathcal{X} \bigg] 
 \end{align} 
where we used $Z-Z_q^{'}=(X_q-X_q^{'})/t_{\rm max}$, if the maximal path does not change when $X_q^{'}$ is substituted for $X_q$; if on the other hand the maximal path changes then, $Z-Z_q^{'}\le (X_q-X_q^{'})/t_{\rm max}$, which can be verified by considering a substitution of $X_q^{'}$ in the original path which is smaller than $Z_q^{'}$. It is straightforward to show that the RHS of equation~(\ref{eqn:final_var_eq}) is given by ${\sigma^2}/{t_{\rm max}}$, which concludes the proof. \\

In order to bound the variance of our max-consensus algorithm, we use Theorem~\ref{the4:var} to write $x_i(t)$ in the first run of the algorithm with zero initial measurements as,  
\begin{equation}
	x_i(t) = \lambda t + \sigma \sqrt{t} Y_t,
\end{equation}
where $Y_t$ is an auxiliary random variable with ${\rm Var}(Y_t)\le 1$, which is clearly equivalent to Theorem~\ref{the4:var} after using $\hat{{\lambda_i}}(t_{\rm max})=x_i(t)/t$.

In the second run of the algorithm after $D$ iterations, where $D$ is the diameter of the network, all nodes converge on the maximum of the initial measurements. Hence we can write our estimator $\hat{{\lambda_i}}(t_{\rm max})$ as,
\begin{equation}\label{eqn:var_run1}
	x_i(D) = (\lambda - \hat{{\lambda_i}}(t_{\rm max})) D + \sigma \sqrt{D} Y_D + x_{\rm max}(0),
\end{equation}
where we know that,
\begin{equation}\label{eqn:hat_lambda}
	\hat{{\lambda_i}}(t_{\rm max}) = \lambda + \frac{\sigma}{\sqrt{t_{\rm max}}} V_{t_{\rm max}}
\end{equation}
where $V_{t_{\rm max}}$ is an auxiliary random variable with ${\rm Var}(V_{t_{\rm max}}) \le 1$. Since the two runs involve independent noise samples, substituting equation~(\ref{eqn:hat_lambda}) into equation~(\ref{eqn:var_run1}) gives,
\begin{equation}
	{\rm Var}(x_i(D)) \le \sigma^2 \bigg(\frac{D^2}{t_{\rm max}} + D\bigg).
\end{equation}
This shows that the variance of our estimator scales linearly with the diameter of the network, as long as $t_{\rm max}$ also scales linearly with $D$.

\setlength{\textfloatsep}{3pt}
\begin{figure}[h]
	\centering
	\includegraphics[width=0.4\textwidth]{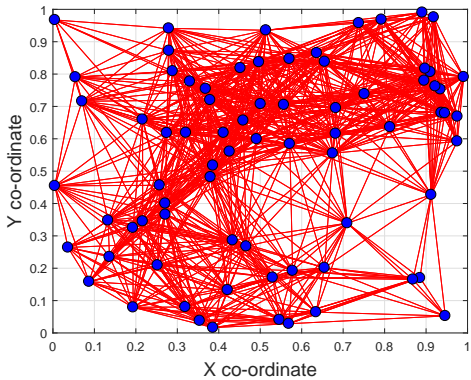} 
	\caption{Network with $N=75$ nodes.} 
	\label{fig1:network}
\end{figure}
\setlength{\textfloatsep}{3pt}
\section{Simulation Results}
\label{sec8:simulations}
We consider a distributed network with $N=75$ nodes, as shown in Figure \ref{fig1:network}. This irregular graph was randomly generated, which is commonly followed \cite{a2_nejad2009max,a4_zhang2016max,aa1_nejad2010max,r1_improved_maxbounds}.
The spectral radius of the graph generated was computed to be $\rho = 30.56$. 
We consider two graph topologies for the simulations:
\begin{enumerate}[i)]
	\item Fixed graphs : by selecting $p=0$ as in Figure \ref{fig1:network}.
	\item Time-varying graphs (Random graphs) : by selecting $p=0.5$.
\end{enumerate}
Communication links between any two nodes has a noise component distributed as $\mathcal{N}(0,1)$.  
First, all nodes are initialized to $0$ and the max consensus algorithm is run to estimate growth rate $\hat{{\lambda_i}}(t_{\rm max})$ as in line \ref{alg:rate est} of the algorithm. Note that, following results are Monte-Carlo averaged over $500$ iterations.

\setlength{\textfloatsep}{2pt}
\begin{figure}[h]
	\centering
	\includegraphics[width=0.43\textwidth]{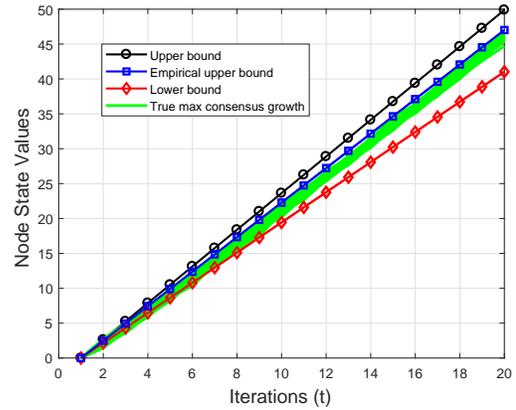} 
	\caption{Comparison of upper bound, lower bound and the max update from equation~$(1)$ for all nodes with $\mathcal{N}(0,1)$ additive noise for a fixed graph with $N=75$.} 
	\label{fig:upper_bound}
\end{figure}
\setlength{\textfloatsep}{2pt}
\begin{figure}[h]
	\centering
	\includegraphics[width=0.43\textwidth]{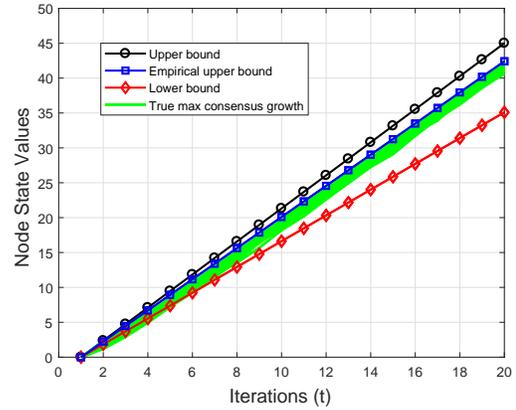} 
	\caption{Random graphs with $N=75$ and edge deletion probability of $p=0.5$.} 
	\label{fig:upper_bound_rand}
\end{figure}

\subsection{Efficiency of the bounds}
For fixed graphs, we compare the upper bound given by equation~(\ref{eqn:ub_gauss}), empirical upper bound, lower bound given by equation~(\ref{eqn:lowb_irreg}), and the Monte-Carlo estimate of max consensus growth is plotted for every node and labeled as ``True max-consensus growth'' in Figure \ref{fig:upper_bound}. We observe in Figure \ref{fig:upper_bound} that the empirical upper bound in Section~\ref{sec:empirical_ub} is much tighter than the original upper bound. 

The same experiment was repeated on a random graph, which was obtained by randomly deleting each edge of the graph in Figure~\ref{fig1:network} with probability $p=0.5$. The comparison of the upper bound, given by equation~(\ref{eqn:sup_rand}), empirical upper bound, lower bound given by equation~(\ref{eqn:lowb_rand}), and the true Monte-Carlo estimate of the max consensus growth is shown in Figure~\ref{fig:upper_bound_rand}. Note that, not only the empirical upper bound is tight for time-varying graphs, but it is also generalizable for different graph topologies. 

\setlength{\textfloatsep}{2pt}
\begin{figure}[h]
	\centering
	\includegraphics[width=0.45\textwidth]{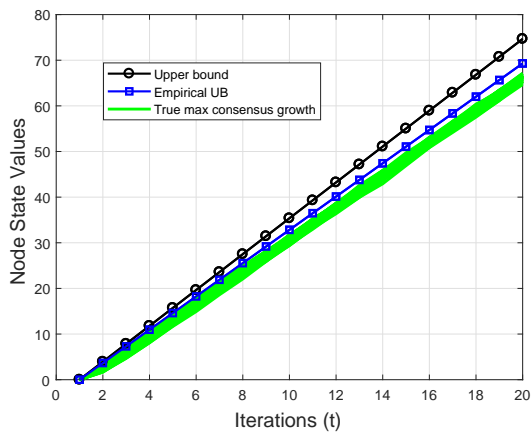} 
	\caption{Comparison of upper bound, empirical upper bound and max consensus growth rate for a network in Figure~\ref{fig1:network} with $N=75$ and $p=0$, where the noise on the links are sampled from Laplace distribution with zero mean and unit variance.} 
	\label{fig:lap_noise}
\end{figure}
\setlength{\textfloatsep}{3pt}
\begin{figure}[hbt]
	\centering
	\includegraphics[width=0.45\textwidth]{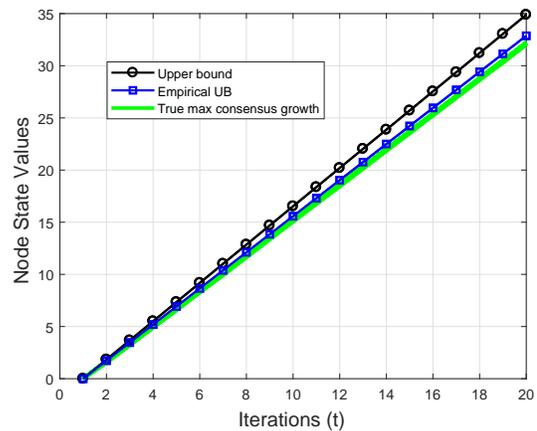} 
	\caption{Comparison of Upper bound, empirical upper bound and max consensus growth rate for a network in Figure~\ref{fig1:network} with $N=75$ and $p=0$, where the noise on the links are sampled from uniform distribution with zero mean and unit variance.} 
	\label{fig:unif_noise}
\end{figure}

Next, we run simulations for non-Gaussian distributions such as Laplace and Uniform distributions to verify the tightness of upper bound.
In Figures~\ref{fig:lap_noise}-\ref{fig:unif_noise}, we compare the performance of upper bound and empirical upper bound for network in Figure~\ref{fig1:network} with $N=75$, where the noise on the links are sampled from Laplace and continuous uniform distributions, respectively. The parameters of Laplace distribution $L(\mu,b)$ were chosen as $\mu=0$ and $b=1/\sqrt{2}$, and uniform distribution $U(a,b)$ as $U(-\sqrt{3},\sqrt{3})$, to ensure zero mean and unit variance. Results also show that the empirical upper bound holds good for general noise distributions. Since Laplace distribution is heavy-tailed compared to Gaussian and uniform, it has a larger growth rate.

\subsection{Performance of the algorithms}

We compare the performance of conventional max consensus algorithms and the proposed algorithm, subjected to additive Gaussian noise $\mathcal{N}(0,1)$.
In order to represent the actual sensor measurements, for both fixed and random graphs, we consider a synthetic dataset with nodes initialized with values over $(100,200)$, where the true maximum of the initial state values is $200$. The robust max consensus algorithm given in Algorithm~$1$ is run over these initial measurements on both the graphs. The results are Monte-Carlo averaged over $500$ iterations.

\begin{figure}[!htb]
	\centering
	\includegraphics[width=0.43\textwidth]{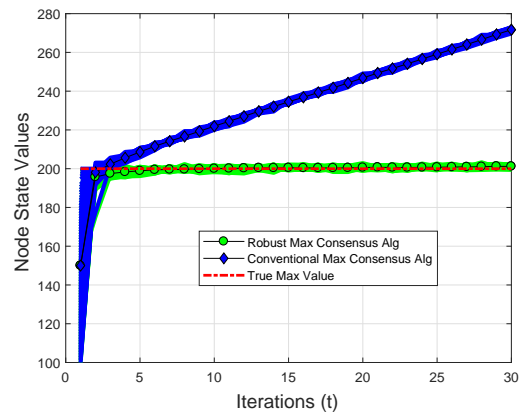} 
	\caption{Performance of the proposed algorithm in the presence of additive noise from $\mathcal{N}(0,1)$ for fixed graphs.} 
	\label{fig3:compare both}
\end{figure}

\begin{figure}[!htb]
	\centering
	\includegraphics[width=0.43\textwidth]{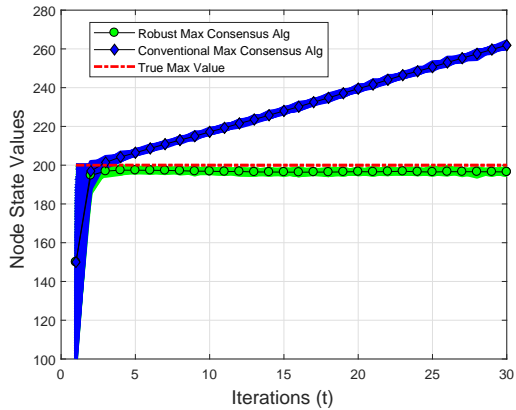} 
	\caption{Performance of the proposed algorithm in the presence of additive noise from $\mathcal{N}(0,1)$ for random graphs with probability of edge deletion $p=0.5$.} 
	\label{fig3:compare both_rand}
\end{figure}

For fixed graphs, performance of our robust max consensus algorithm and the existing max based consensus algorithm is shown in Figure \ref{fig3:compare both}. It can be observed that the conventional max consensus algorithm diverges as $t$ increases, whereas our algorithm does not suffer from increasing linear bias. Even in case of random graphs, our algorithm converges to the true maximum, whereas the conventional max consensus algorithm diverges as $t$ increases, as shown in Figure~\ref{fig3:compare both_rand}.

\setlength{\textfloatsep}{2pt}
\setlength{\textfloatsep}{2pt}
\begin{figure}[h]
	\centering
	\includegraphics[width=0.43\textwidth]{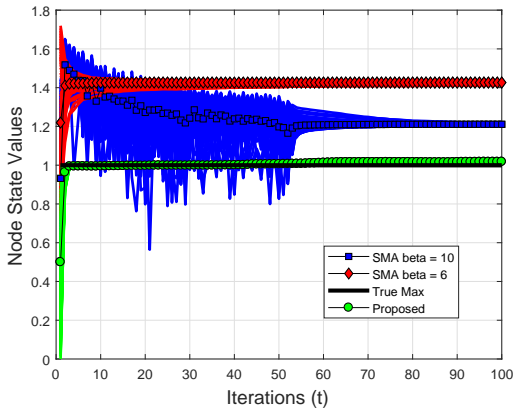} 
	\caption{Comparison of the proposed algorithm and soft-max based average consensus algorithm (SMA) \cite{a4_zhang2016max}, for a graph with $N=75$, $\beta$ for SMA as $\{6,10\}$ in the presence of additive noise $\mathcal{N}(0,1)$, distributed over the edges.
} 
	\label{fig:sai_vs_rma}
\end{figure}
\setlength{\textfloatsep}{3pt}

By comparing the dynamic range of growth rate of conventional max consensus algorithms in Figure~\ref{fig3:compare both} and Figure~\ref{fig3:compare both_rand}, we observe that a) at $t=30$, state values over fixed graphs has mean and standard deviation of $270.39$ and $0.6966$ respectively, and b) at $t=30$, state values over random graphs with $p=0.5$ has mean and standard deviation of $261.09$ and $0.9233$, respectively. Thus, node state values grow slower for random graphs with $0 < p < 1$, compared to fixed graphs $(p=0)$ due to the reduced connectivity of the graph. 

\subsection{Comparison with existing works}
The performance of our proposed algorithm was compared with the conventional max consensus algorithm \cite{a1_iutzeler2012analysis,a2_nejad2009max,a3_giannini2016asynchronous} in Figures~\ref{fig3:compare both}-\ref{fig3:compare both_rand} and clearly, conventional max consensus algorithm diverges in the presence of additive noise. 

Additionally, we compared the performance against the soft-max based average consensus algorithm (SMA), proposed in \cite{a4_zhang2016max}, as shown in Figure~\ref{fig:sai_vs_rma}. The soft maximum of a vector $\mathbf{x} = [x_1,\cdots,x_N]$ is denoted as:
$$
{\rm smax(\mathbf{x})}=\frac{1}{\beta}\log\sum_{i=1}^{N} e^{\beta x_i},
$$
where $\beta > 0$ is a design parameter.
 We consider the same network with $N=75$ as in Figure~\ref{fig1:network}. Nodes were initialized linearly over $(0,1)$. We considered the design parameter $\beta$ of the SMA algorithm to be $\beta = \{6,10\}$. The proposed algorithm and the SMA algorithm were applied in the presence of additive noise $\mathcal{N}(0,1)$, distributed over the edges. 

The SMA algorithm with $\beta=6$ converges faster than with $\beta = 10$, however, $\beta=6$ has greater estimation error than $\beta = 10$. In comparison with SMA, our proposed algorithm performs better in terms of bias and variance of the estimate of true maximum value, and the number of iterations required for convergence.

\section{Conclusion}
\label{sec9:conclusion}
A practical approach for reliable estimation of maximum of the initial state values of nodes in a distributed network, in the presence of additive noise is proposed. Firstly, we showed the existence of a constant growth rate due to additive noise and then derived upper and lower bounds for the growth rate.
It is argued that the growth rate is constant, and the upper bound is a function of spectral radius of the graph. By deriving a lower bound, we proved that the growth rate is always a positive non-zero real value. We also derived upper and lower bounds on the growth rate for random time-varying graphs. An empirical upper bound is obtained by scaling the original bound, which is shown to be tighter and generalizable to different networks and noise settings. Finally, we presented a fast max-based consensus algorithm, which is robust to additive noise and showed that the variance of the growth rate estimator used in this algorithm decreases as $\mathcal{O}(t_{\rm max}^{-1})$ using concentration inequalities. We also showed that the variance of our estimator scales linearly with the diameter of the network. Simulation results corroborating the theory were also provided.

\bibliographystyle{IEEEtran}
\bibliography{TSIPONreferences}
\end{document}